*Accepted to the Low Temperature Physics, Revised version (Invited)*

# Impact of Carbon Contamination on the Low-Temperature Electric Conduction of the Spark-Plasma Sintered Barium Titanate Ceramics

Oleksandr S. Pylypchuk[1], Taisiia O. Kuzmenko[1,3], Denis O. Stetsenko[1], Oleksii V. Bereznykov[1], Serhii E. Ivanchenko[2], Ihor M. Danylenko[3], Anna N. Morozovska[1], Volodymyr M. Poroshin[1], Victor V. Vainberg[1, *]

*[1]Institute of Physics of the National Academy of Sciences of Ukraine, 46 Nauky Avenue, 03028 Kyiv, Ukraine*

*[2]Frantsevich Institute for Problems in Materials Science, National Academy of Sciences of Ukraine, Omeliana Pritsaka str., 3, Kyiv 03142, Ukraine*

*[3]V. Lashkarev Institute of Semiconductor Physics, National Academy of Sciences of Ukraine, 41, Nauky Avenue, 03028 Kyiv, Ukraine*

Oleksandr S. Pylypchuk 0000-0003-0136-0799

Taisiia O. Kuzmenko 0009-0002-4232-8488

Denis O. Stetsenko 0009-0002-9141-0939

Oleksii V. Bereznykov 0000-0001-6394-7304

Serhii E. Ivanchenko 0000-0002-9859-1835

Anna N. Morozovska 0000-0002-8505-458X

Ihor M. Danylenko 0000-0002-8740-204X

Volodymyr M. Poroshin 0000-0001-8217-3949

Victor V. Vainberg 0000-0002-9840-8033

---

[*] corresponding author, e-mail: viktor.vainberg@gmail.com

## Abstract

The dielectric and electric conduction properties of $BaTiO_3$ samples fabricated by the spark plasma sintering with additional contamination by different content of carbon have been investigated in the temperature range from 77 through 408 K in small non-heating electric fields and in the range of 77-200K under strong electric fields up to 20 kV/cm. The effective dielectric permittivity of the samples with additional carbon achieves colossal values at low frequency, up to several million at 393 K. In the low temperature range, it remains very high magnitude, $10^4$ – $10^5$ at low frequency, and is strongly dependent on the carbon content. The electric conduction in the range of 77 – 200 K bears the hopping conduction features and obeys the Mott law. We also studied the electric conduction and effective dielectric permittivity vs electric field strength in the range of 77 – 200 K. The results are explained within the frames of hopping conduction theory and suggestion on different constituents in polarization processes and electric conduction.

## 1. Introduction

The barium titanate ($BaTiO_3$) is a classical ferroelectric material with a high dielectric permittivity [1] that is widely used in ceramic capacitors, posistors, pyroelectric detectors, cells of dynamic random-access memories. Recently research investigations focused on the $BaTiO_3$ fabricated by the Spark Plasma Sintering (SPS) method, first proposed in Ref. [2]. The $BaTiO_3$ ceramic obtained by this method possesses noticeably higher dielectric permittivity [3, 4, 5, 6, 7].

The colossal dielectric permittivity of the SPS $BaTiO_3$ ceramics observed in Refs. [5-6] was explained by the complex interplay between a possible diffuse ferroelectric-paraelectric phase transition and the Maxwell-Wagner (MW) effects, which emerge from the formation of spatial charges at interfaces between different materials, and in the ferroelectric nanoparticle-air interface [3, 8, 9]. Notably that the MW-type effective dielectric permittivity reaches colossal values (more than $10^5$) at low frequencies. At low frequencies, microscopic inhomogeneities in electrical conductivity (mainly between grains/particles and their boundaries) give rise to interfacial charge accumulation, producing MW-type polarization and internal barrier layer capacitance (IBLC), resulting in the experimentally observed colossal effective permittivity in nanograined ferroelectric ceramics, nanopowders, and/or nanocomposites [10]. Additional contributions to the colossal permittivity may arise from inhomogeneous layers between the electrodes and the sample, known as surface barrier layer capacitance (SBLC) [3, 8-10]. These interfacial polarization mechanisms can be described by the effective medium approach (EMA) [11], as discussed in Refs. [12, 13], and in the works of Petzelt et al. [14] and Rychetský et al. [15]. Both the IBLC and SBLC effects produce a colossal dielectric response at low frequencies only. For the systems where enhancement of the dielectric permittivity due to the ferroelectric-paraelectric transition dominates, the high dielectric response is nearly frequency-independent up to the MHz frequency range. In theory, contributions of the MW-type polarization and the ferroelectric polarization to the dielectric response may be separated by the broadband dielectric spectroscopy methods [3], but they are strongly intertwined and become inseparable in the nanostructured systems with the diffuse relaxor-like ferroelectric phase transitions. The main reasons for the phase transition diffuseness are the wide distribution of nanoparticle/grain sizes, strongly inhomogeneous elastic strains and compositional disorder (e.g., chemical strains) [3-10].

It is important that the technology of SPS ceramic fabrication results in some contamination by carbon, which further one tries to eliminate as much as possible. The advances of the SPS $BaTiO_3$ as compared to other traditional techniques, for example hot-pressing sintering, are well studied and described, for example, in Refs. [3 – 6, 16, 17, 18] and references herein. This fabrication technique results, as a rule, in the colossal magnitude of the effective dielectric permittivity, which physical origin is unclear.

The SPS $BaTiO_3$ is in fact a granular material consisting of sintered nanosized particles conditionally close to spherical shape with overall dimensions of several tens through several hundred nm. Therefore, they usually are considered as agglomerates of grains, which cores, surface shells and intergrain space possess own different mechanisms determining both polarization and electric transport properties. The single crystalline $BaTiO_3$ possesses 4 different crystallographic forms in the wide temperature range: rhombohedral below ~190 K, orthorhombic between 190 K and ~272 K, tetragonal between 272 K and 393 K, and cubic above 393 K [19, 20]. Electronic transport both in the conduction band and via the polaron hopping by localized states – all these forms are more or less studied (see for example, Refs. [21, 22, 23, 24]). In the case of SPS $BaTiO_3$ ceramics the main attention is focused on the dielectric permittivity and analysis of mechanisms determining its constituents. Meanwhile, the barium titanate contaminated by carbon manifests noticeable electric conduction that is determined by the hopping conduction mechanisms [6]. Also, the effective dielectric permittivity, that is usually relatively small at low temperatures, increases by orders of magnitude manifesting along the conductivity certain dependence on added carbon amount.

In this work we carried out a purposeful study of low temperature electric conductivity of the SPS $BaTiO_3$ ceramics with additionally added carbon to clarify the character both of low temperature conductivity and dielectric response in such specific ferroelectric material in dependence on the carbon amount and reveal interrelation between changes in permittivity and conductivity. For this purpose, along with customary conductivity vs temperature measurements, we performed the measurements of conductivity vs electric field in a much wider field range and transient current vs time waveforms characterizing polarization processes. The results are analyzed within the frames of known theory of hopping conduction, and the observed discrepancy is discussed.

## 2. Experimental

The synthesis procedure implemented to fabricate the samples studied is the same as described in Ref. [6]. The overall dimensions of the $BaTiO_3$ nanoparticles were 24 nm in average. The sintering temperature was 1100 $^{\circ}$C under uniaxial pressure of 50 MPa, duration about 5 min. The carbon content in the initial mixture varied from 0 to maximum 1 %. We separate 3 groups of samples with noticeable difference in resistivity, especially at low temperatures, and denote them as S#1, S#2 and S#3, according to increasing their resistivity. Also, we used the samples S#4 without any intentionally added carbon (see T**able 1** in **Appendix**). The comparative characterization of the carbon distribution/content in the samples was performed as described in Ref. [6]. It showed that the highest carbon concentration corresponds to the S#1 sample and the lowest one is in the S#3 sample. Also, there were made samples S#2.1 and S#3.1 in the form of thin plates for measurements of current-voltage characteristics under strong electric fields. The thin plates S#2.1 and S#3.1 were cut from the large pellets S#2 and S#3, respectively.

Analysis of the Raman spectra for final sintered samples S#1, S#2 and S#3 showed that carbon distribution is rather inhomogeneous; it has the form of clusters with the effective lateral correlation length of ordered *sp*2-carbon regions, further referred as "effective cluster size", comparable with size of the barium titanate nanoparticle (~24 nm) with deviation of ±6 nm. Let us underline that the effective lateral correlation length is related but not equal to the geometrical sizes of carbon clusters.

The effective sizes and spread of clusters lateral correlation length in the samples S#1, S#2 and S#3 are illustrated in **Fig. 1**. The filled symbols denote the effective cluster size determined at chosen points on the sample.

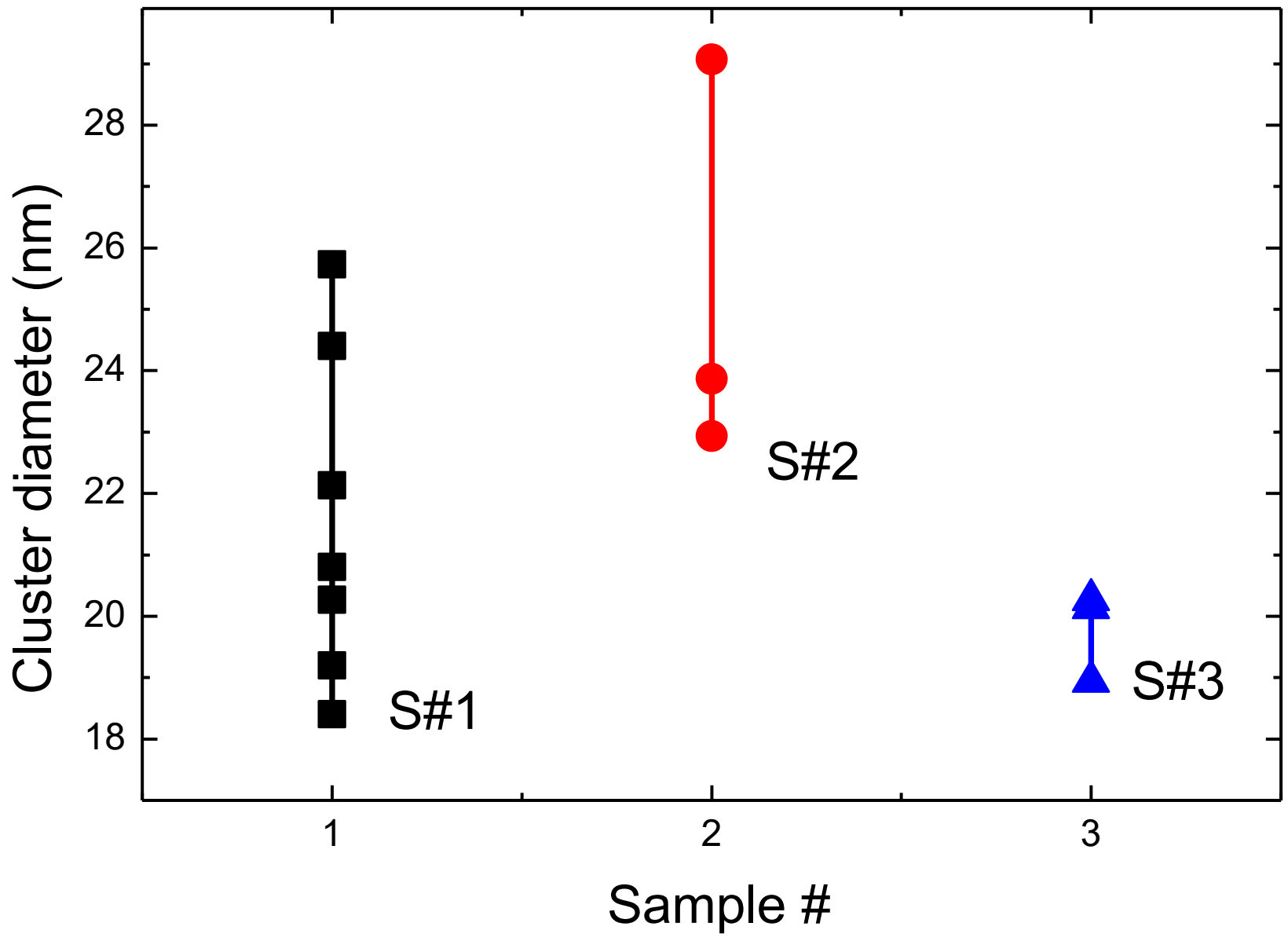

**Figure 1.** Effective cluster size (lateral correlation length) of the carbon inclusions in the $BaTiO_3$ matrices in three kinds of studied samples.

Such picture enables us to suggest the carbon to reside on the $BaTiO_3$ grain surface more or less covering them.

The measurements of electric conduction and dielectric response were performed using the samples shaped as a parallelepiped with dimensions listed in **Table 1** in Appendix. The electric contacts were made by thermal deposition of Indium. The measurements of dielectric permittivity and electric conduction were performed in the temperature range of 77 through 300 K in a cryostat, and 300 through 408 K in a thermostat. The dielectric permittivity and resistivity in the AC regime were determined from the capacitance and resistance measured by the LCR-meter LCX200 ROHDE & SCHWARZ. These measurements were performed in the frequency range from 4 Hz through 500 kHz. Measurement of the DC resistivity was performed in the circuit consisted of a sample serially connected with a load resistor (10 Ohm). The voltage was applied by a programmable DC power supply GW Instek PSP-603. The voltage across the sample and load resistor was measured by the digital multimeters Keithley-2000. The current was determined from the voltage across the load resistor.

Measurements of the dielectric permittivity and electric conduction characteristics under strong electric fields were performed in a similar circuit in the pulsed voltage regime with a voltage pulse being sufficiently short (no longer 14 μs) to avoid the Joule heating of samples. The cable capacitance of the load resistor link is ≈ 100 pF. The waveform of the voltage across the load resistor has an exponential decay shape. It may be fitted by the function $y_0 +$ $y_1 exp[-(t - t_0)/\tau]$. From $y_0$ we obtain the DC current magnitude to determine the DC resistivity $\rho$, from $\tau = \varepsilon_0 \varepsilon \rho$, where $\varepsilon_0$ is the vacuum dielectric permittivity, we obtain the effective permittivity value. The resistance of the load resistor is sufficiently small in order not to distort measured parameters. As shown below the decay curves for the samples studied are well fitted by the two-exponential decay which enable us to separate two constituents of polarization processes.

## 3. Results and discussion

Shown in **Fig. 2a** are temperature dependences of dielectric permittivity measured at frequency of 10 Hz in the low-voltage non-heating regime. The red dashed lines separate the temperature ranges of the crystallographic forms characteristic for the bulk single crystalline $BaTiO_3$. The direct correlation with our granular material so far is not possible, and the bulk temperatures must be considered as a reference markers only. Assuming that finite size effects, which can modify these phase transition temperatures, are not important for some reasons, one may assume which phase in what range prevails. However, it is well-established that finite size and strain effects can be principally important for small 24 nm $BaTiO_3$ nanoparticles, they can change phase transition temperatures to hundred degrees due to the depolarization field influence, screening of the uncompensated bound charges by (external or/and internal) free carriers, domain structure formation and strain-electrostriction coupling (see e.g., Ref. [25] and refs. therein). The depolarization field always decreases the ferroelectric-paraelectric transition temperature, while its screening by free carries and vortex-like domain formation supports the ferroelectric states. The influence of elastic strains is dual: compressive strains suppress or destroy the ferroelectricity in spherical nanoparticles and thus decrease the ferroelectric-paraelectric transition temperature, while tensile strains, as a rule support it, and increase the ferroelectric-paraelectric transition temperature. At the same time, the size and strain effects become much less significant under the particle size increase above 100 nm. Assuming that a significant amount of small 24-nm nanoparticles is sintered into much larger agglomerations during the SPS, we can consider the SPS ceramic samples as a mixture of microscale $BaTiO_3$ agglomerations and small $BaTiO_3$ nanoparticles, which are separated (at least partially) by carbon inclusions. The relative fractions of the agglomerations (with bulk-like ferroelectric properties) and nanoparticles (with size-driven polar properties) are unknown a priori; it may depend on carbon content. The fraction of carbon is known, but the morphology of its inclusions can vary in the studied samples.

**Figure 2b** shows dependences of resistivity vs temperature, corresponding to these samples. The empty symbols correspond to the DC conduction regime, and the filled symbols – to the AC conduction regime at frequency $f = 10$ Hz. The vertical red dashed lines here also separate the ranges of the crystallographic forms corresponding to single crystalline $BaTiO_3$. This separation enables us to understand to some extent changes in the character, both the dielectric permittivity and electric conduction in the microscale $BaTiO_3$ agglomerations. In the high temperature range of the tetragonal phase near to the range of the cubic phase the dielectric permittivity tends to shape a maximum, though the sample S#4 intentionally undoped seems not to form any hint on maximum. Due to the absence of carbon inclusions, which may act as "spacers" and "screeners" for individual nanoparticles, we may expect the appearance of a very diffuse relaxor-like paraelectric-ferroelectric transition in the sample S#4, whose very broad maximum may be located at higher temperatures outside the measured temperature range. The resistivity vs temperature dependence in this range noticeably increases its slope and accelerates decreasing of resistivity with growing temperature. In two low temperature ranges the effective dielectric permittivity decreases more or less strongly with lowering temperature. Note that the dielectric permittivity in the samples studied is very large, achieving several million in the tetragonal region. And it is obviously dependent on the carbon content, showing at that a reciprocal relation of permittivity and resistivity.

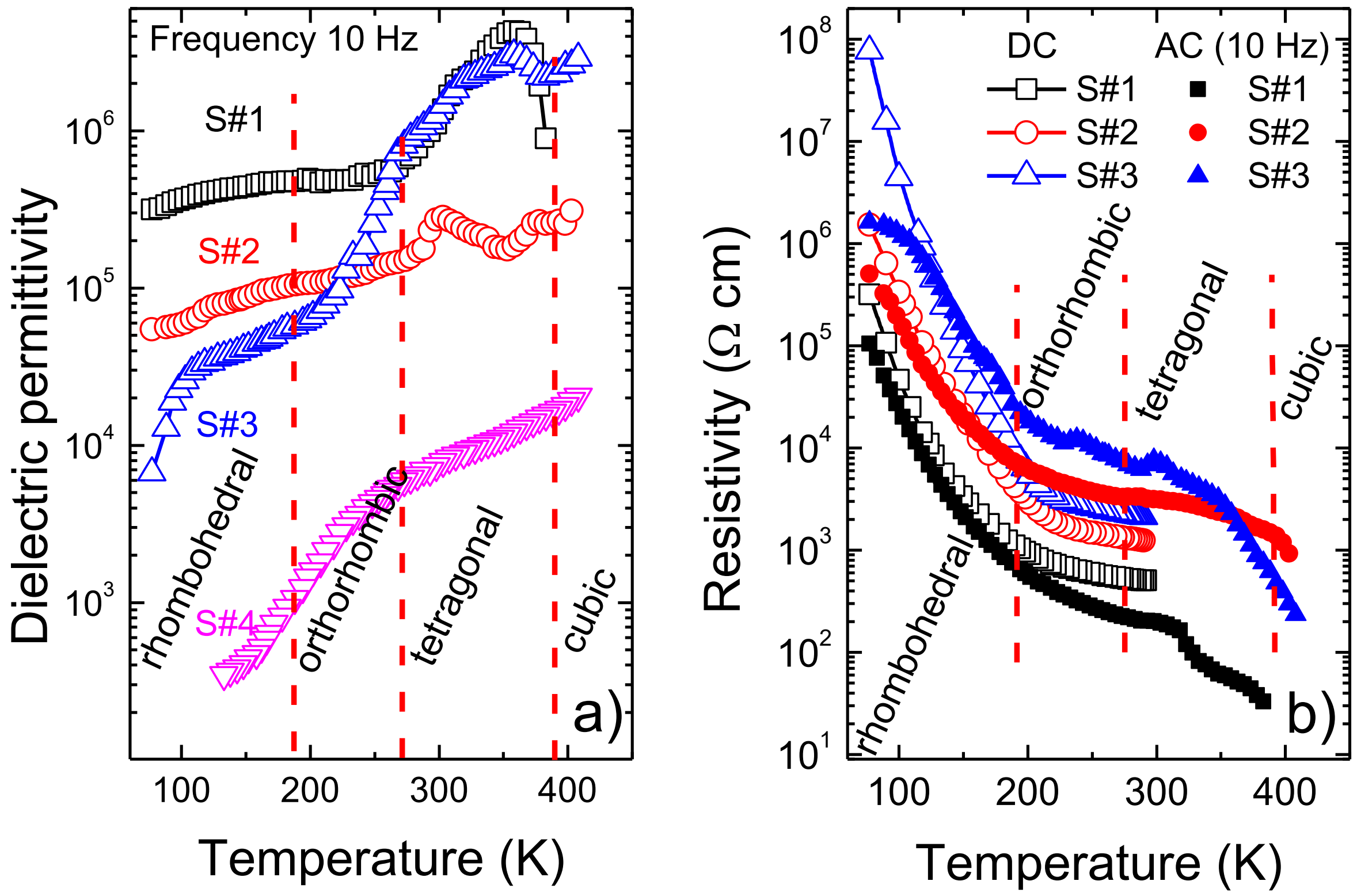

**Figure 2. (a)** Temperature dependences of the dielectric permittivity measured at the frequency of 10 Hz. The red dashed lines separate temperature ranges of different crystallographic forms of the single crystalline $BaTiO_3$. **(b)** Temperature dependences of the DC and AC (10 Hz) resistivity. The red dashed

lines separate the same ranges as in the panel (a). The empty symbols correspond to the DC conduction regime, and the filled symbols – to the AC conduction regime at $f = 10$ Hz.

With lowering temperature all samples demonstrate a strong decrease in the magnitude of permittivity. In the temperature range corresponding to the rhombohedral phase in a bulk single crystalline $BaTiO_3$ the DC resistivity vs temperature dependence, as shown in **Fig. 3**, very well obeys the Mott law for the variable range hopping conduction [26]

$$\rho \sim \exp\left(\frac{T_0}{T}\right)^{1/4}, \qquad T_0 = \frac{21}{k_B N(E_F) a^3}, \tag{1}$$

where $N(E_F)$ is the density of localized electron states in the vicinity of the Fermi level, $a$ is the localization radius of a charge carrier wave function. The parameter $T_0$ for the curves shown in **Fig. 3** is $6.56 \cdot 10^7$ K, $7.17 \cdot 10^7$ K and $3.32 \cdot 10^8$ K for the samples S#1, S#2 and S#3, respectively. The thermo-emf at the room temperature for all three samples S#1, S#2 and S#3 has a sign corresponding to the electron kind conduction (with the positive voltage sign at the "hot" contact relative to the "cold" one) and is about 320 μV/K. According to Refs. [21, 22, 27] the n-type conduction in a single crystalline $BaTiO_3$ is usually determined by small-polaron hopping mechanism. In the case of SPS $BaTiO_3$ ceramics the intergrain ionic transport should be also taken into account along with other mechanisms (including small polarons). Its contribution depends obviously on the carbon content and may impact on the Mott variable-range hopping (VRH) conduction because of the fact that the ionic defects and intergrain structural disorder may create or modify localized electronic states and intergrain barriers. The ion transport seems also to impact on the low frequency dielectric permittivity leading to extremely large effective permittivity, especially at high temperatures.

The dielectric permittivity in the low temperature range varies with temperature much slower than at higher temperatures, while remaining a relatively large magnitude. In the AC regime the conductivity also obeys the law with the activation energy lowering with decreasing temperature with a tendency to saturate resistivity at the lowest temperature (compare corresponding curves with empty and filled symbols in **Fig. 2**).

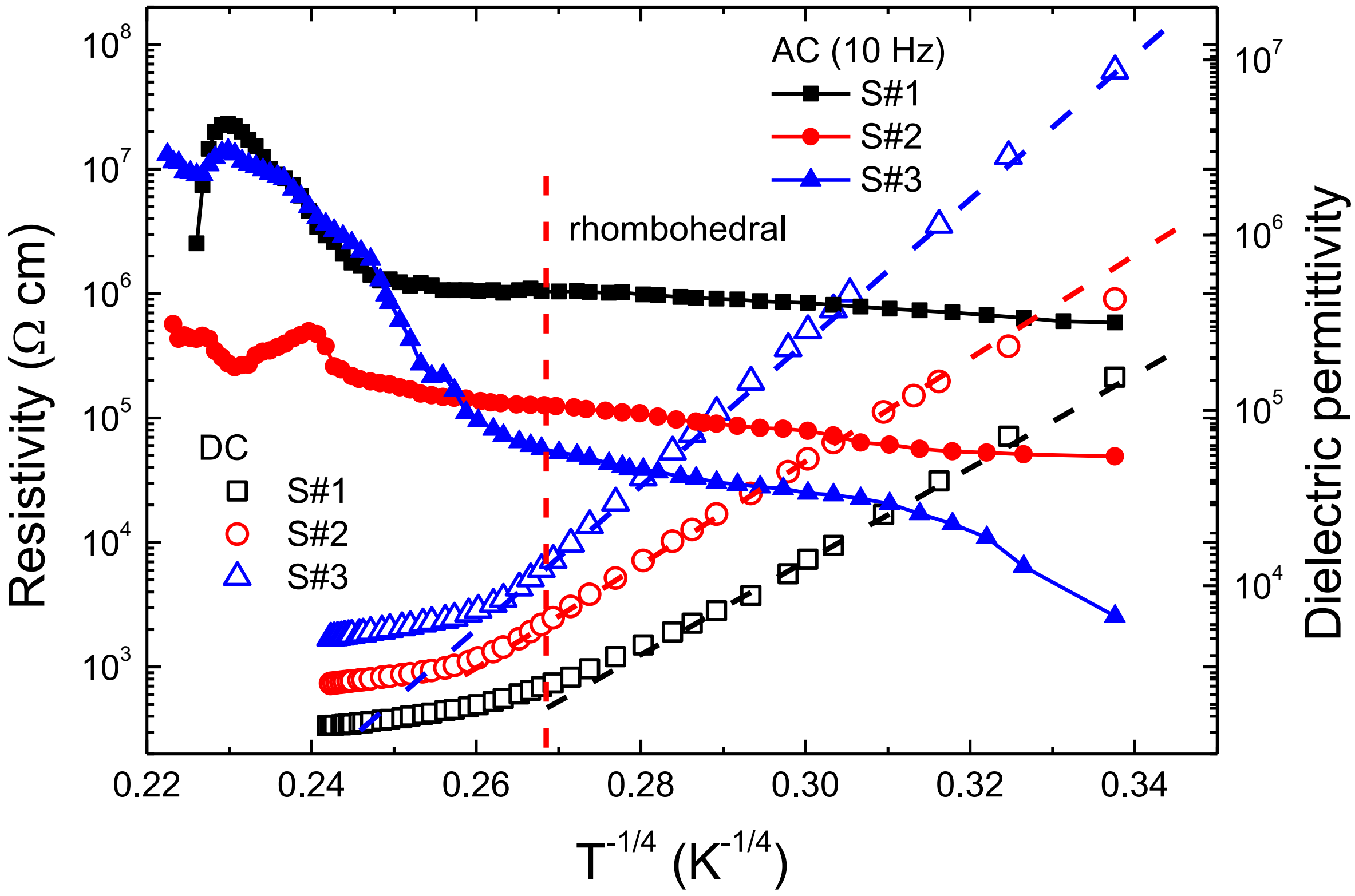

**Figure 3.** Comparison of the DC resistivity (empty symbols) and dielectric permittivity (filled symbols) vs temperature dependences in the Mott law coordinates for 3 samples with different percentages of the carbon content.

As seen from **Figs. 2** and **3**, the more carbon content, the less resistivity and the larger permittivity in the low temperature ranges. Moreover, the more carbon content, the less the resistivity vs temperature slope in the Mott law dependence. This indicates the direct relation between the carbon content and effective permittivity in the low temperature range. The magnitude of permittivity along with impurity concentration is a determining factor for the hopping conduction mechanism. At the same time, it remains unclear how variation of the dielectric permittivity with temperature impacts on the resistivity vs temperature dependence according to the formula (1) since, at first glance, the variation of ε with lowering temperature must change also the parameter $T_0$. Moreover, the changes have self-consistent behavior, because the influence of the screening carriers on the "bare" (i.e., initially unscreened) uncompensated bound charge emerging from the spontaneous polarization determines the polarization value in small nanoparticles [25]. In other words, the density $N(E_F)$ should depend (more or less strongly) on the local band bending induced by the uncompensated internal depolarization field. Once the spontaneous polarization of the $BaTiO_3$ nanoparticle becomes self-screened by the localized carriers the depolarization field drops down and the band bending decreases. The equilibrium barrier height is determined by the complex interplay between the positive depolarization field energy, negative Landau-type polarization energy of the nanoparticle core, the electrostatic energy of screening charges (localized mainly in the shell). Incomplete screening may stimulate the domain formation that contributes to correlation, electrostatic and elastic energies. All these contributions are time-dependent, and their changes determine the frequency relaxation spectra of the inhomogeneous system in applied electric fields [28].

Consideration of the hopping conduction in the EMA implies the localization radius, determining the tunneling probability between localized centers, to be (in SI units)

$$a = \frac{4\pi\varepsilon_0\varepsilon\hbar^2}{m_0 m^* e^2}, \qquad (2)$$

where $m_0$ is the free electron mass, $\hbar$ is the reduced Planck constant, $e$ is the electron charge, $\varepsilon$ is the effective dielectric permittivity, $\varepsilon_0$ is the vacuum permittivity, $m^*$ is the electron effective mass. It suggests the localization radius should also vary with temperature.

Shown below in **Fig. 4** are the dependences of electric conductivity vs electric field strength at several constant temperatures in the lowest temperature range for the S#2.1 and S#3.1 samples.

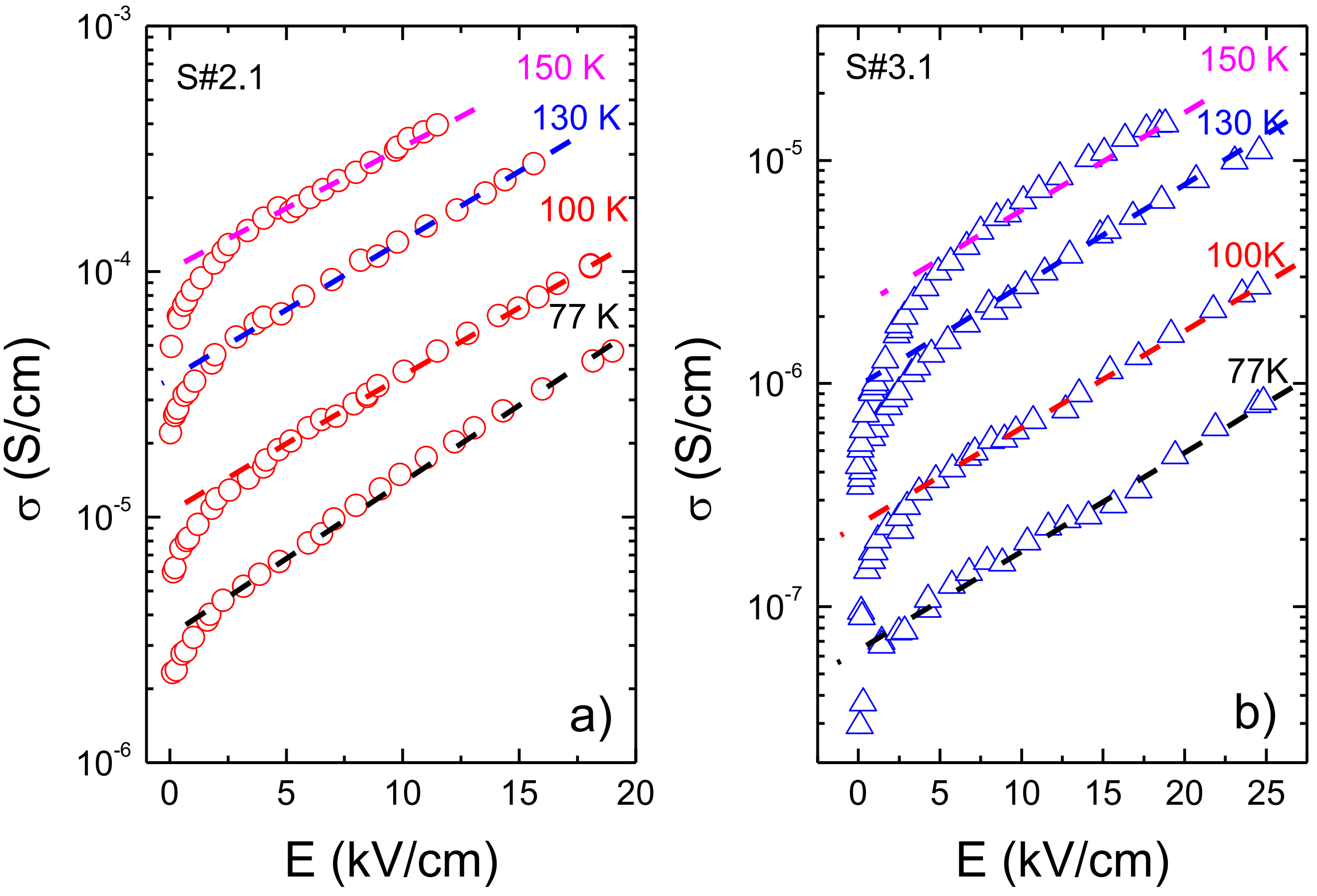

**Figure 4.** Electric conductivity vs electric field strength dependences for the samples S#2.1 and S#3.1 at several constant temperatures in the temperature range of 77 – 150 K.

As seen from **Fig. 4**, approximately above electric field of 1 – 5 kV/cm all curves may be well fitted by straight lines in the coordinates lnσ - $E$, i.e., lnσ is proportional to the electric field strength. According to Refs. [29, 30], under applied electric fields satisfying the condition $eEr_m > k_B T$, the hopping conductivity obeys the law:

$$\sigma(E) \sim \exp\left(C \frac{eEr_m}{k_B T}\right), \qquad r_m = a\xi/2, \quad \xi = \left(\frac{T_0}{T}\right)^{1/4}, \tag{3}$$

where $E$ is the strength of applied electric field, $r_m$ is the maximal hopping length along the percolation path, $e$ is the electron charge, $k_B$ is the Boltzmann constant. The constant $C = 0.8$ [29] or $C = 0.17$ [30]. Note that expressions (3) cannot consider rigorously the inhomogeneous internal (i.e., depolarization) fields emerging in the nanograined SPS ceramics. However, since these fields are characterized by zero spatial-average value, expressions (3) can be applied within the EMA.

At the same time under extremely high electric fields $eEa > k_B T$ according to [31] conductivity does not depend on temperature any longer and obeys the law

$$J \sim \exp\left[-\left(\frac{E_0}{E}\right)^{1/4}\right], \qquad E_0 = \alpha \frac{k_B T_0}{ea}, \tag{4}$$

where α~1. Since there is not seen any hint of transition to the dependence (4) we consider the electric field range in **Fig. 4** as moderately strong electric fields satisfying the criterium

$$\frac{2k_B T}{ea}\left(\frac{T}{T_0}\right)^{1/4} < E < \frac{k_B T}{ea}. \tag{5}$$

We may estimate the localization radius from the low boundary of this criterion. For the samples S#1 and S#2 the magnitude of $(T_0)^{1/4}$ is about 90. At 77 K the boundary field is about 1.5 kV/cm. Hence from (5) we obtain $a \geq 2.92\, nm$. As follows from (2), $a(nm) = 5.3 \cdot 10^{-2} \frac{\varepsilon}{m^*}$, where $m^*$ is the effective mass and in the crystalline $BaTiO_3$ it is known to be 6.5 [32]. Hence the low limit ε in Eq.(2) according to these data is ε ≥ 360.

As seen from **Fig. 5** this value is more appropriate for the intentionally undoped sample S#4 in the entire frequency range. The permittivity of the doped samples S#1, S#2 and S#3 approaches this range of values only at the highest frequency. The discrepancy may be explained by all reasons mentioned above, in the discussion of possible dependence of $N(E)$ on $\varepsilon$. To answer the question what reasons (depolarization, screening or correlation energy change) are decisive (among many) deserves a separate study. As a rule, the dielectric permittivity of weakly screened nanoparticles should be much larger and reveal more pronounced frequency dependence than that of well-screened nanoparticles and large agglomerates [25, 28]. In other words, a very large low-frequency dielectric permittivity may arise due to a substantial contribution of the Maxwell–Wagner mechanism related with the internal grain-boundary and/or electrode barrier layers. Therefore, the large low-frequency effective macroscopic permittivity may be not necessarily considered as the dielectric constant that should determine an effective-Bohr-radius used to estimate the electron localization radius.

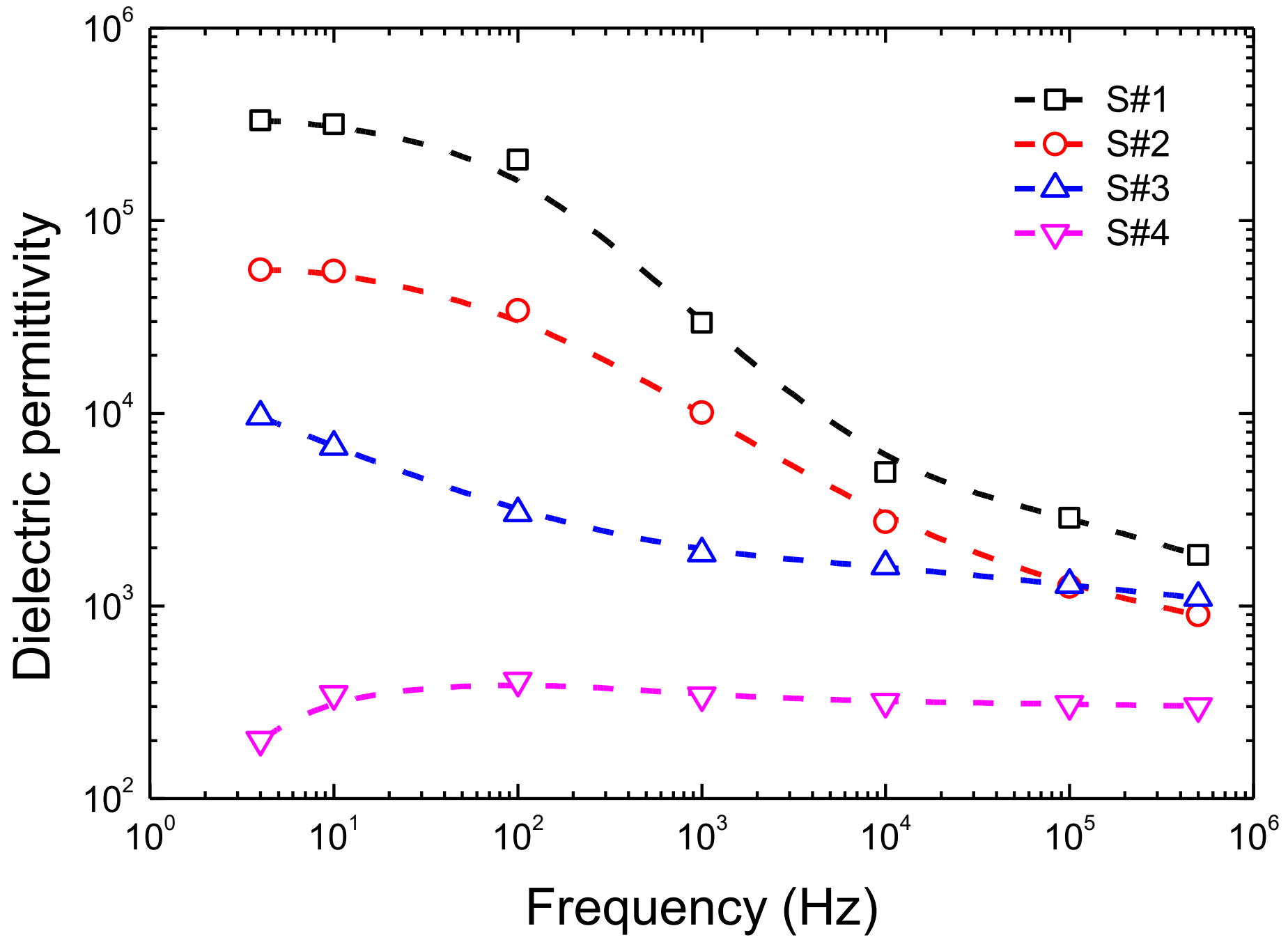

**Figure 5.** Frequency dependences of the dielectric permittivity of the doped samples S#1, S#2 and S#3 and intentionally undoped S#4 at 77 K.

Here we should also note that the slope of the dependence lnσ - $E$ in the field range, where lnσ ~ $E$ very weakly depends on the temperature, is much less than inversely proportional to $T^{5/4}$. This contradicts the formula (3) and leads to suggestion that it may be caused by variation of the dielectric permittivity as well. To elucidate this issue, we determined the dielectric permittivity under the conditions of strong electric field from the decay part of the current through the sample measured in the pulsed regime. Shown in **Fig. 6** are such waveforms measured for the sample S#2.1 at different values of the electric field strength. For the sake of convenience, they are normalized by their maximum at the beginning of pulse. The non-normalized curves for current density can be fitted by

$$J = J_{DC} + \sum_{i=1}^{n} J_i exp\left(-\frac{t}{\tau_i}\right) \cong J_{DC} + J_1 exp\left(-\frac{t}{\tau_1}\right) + J_2 exp\left(-\frac{t}{\tau_2}\right), \quad (6)$$

with 2 exponential decay components, $J_i exp\left(-\frac{t}{\tau_i}\right)$, and one base, $J_{DC}$. Note that the common "base", $J_{DC} \cong J(t \to \infty)$, is required for mathematically correct exponential expansion of the decay curves. The fitting procedure gives a small error for two exponents, $J_1 exp\left(-\frac{t}{\tau_1}\right) + J_2 exp\left(-\frac{t}{\tau_2}\right)$ [6].

Note, the values of $\tau_1$ and $\tau_2$ are of order of hundred ns and μs that is by several orders of magnitude larger than the $RC = 10$ Ohm x $100$ pF $= 1$ ns of the load resistor link used to record the current decay waveforms. Whence, the DC conductivity is $\sigma = \frac{J_{DC}}{E}$ and $\tau = \varepsilon_0 \varepsilon \rho$, where $\varepsilon_0$ is the vacuum dielectric permittivity, $\rho = 1/\sigma$. Whence we obtain two effective permittivity values, namely $\varepsilon_1$ and $\varepsilon_2$. They differ each other approximately by the order of magnitude. We relate this to the relaxation processes inside the core-shell grains and in the intergrain space, which include nanosized $BaTiO_3$ cores covered with shell regions containing localized screening charges, carbon-enriched intergrain regions and bulk-like $BaTiO_3$ agglomerates. We should note that fully unambiguous separation of these processes and relation with physical mechanisms is not so far possible from the present data and demands further comprehensive studies. Therefore, we assign the *apparent* relaxation components to core–shell and intergrain processes as tentative.

The unambiguous analysis of these issues demands further comprehensive investigations and our conclusions on the nature of constituents of the electric transport and dielectric response are based on general concepts of the physical mechanisms characteristic for ferroelectric nanograined materials.

Shown in **Fig. 7** are dependences of both permittivity components vs the electric field strength in comparison with the field dependence of conductivity for the sample S#2.1 at 77 K. One should note that these permittivity components noticeably depend on the electric field strength (**Fig. 7**) and may change with the growing field in the whole range by an order of magnitude. Also note that at $E \approx 15$ kV/cm the conductivity of the sample S#2.1 begins to deviate from the dependence $ln\sigma \sim E$ and above this field the permittivity component $\varepsilon_2$ begins to decrease instead of increasing.

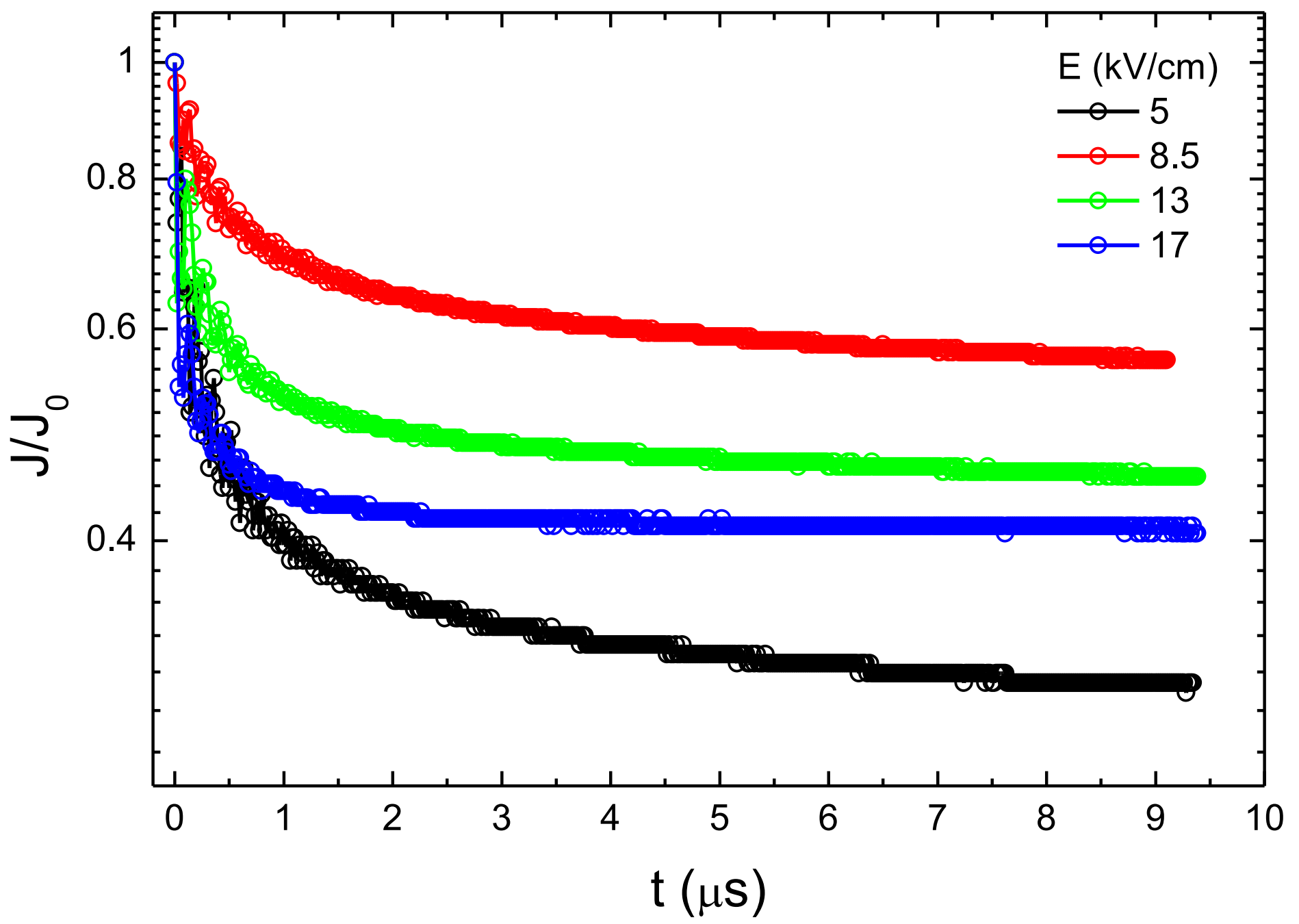

**Figure 6.** Normalized current vs time waveforms of the sample S#2.1 at 77 K for different magnitude of the electric field strength in the field region of the $ln\sigma \sim E$ dependence.

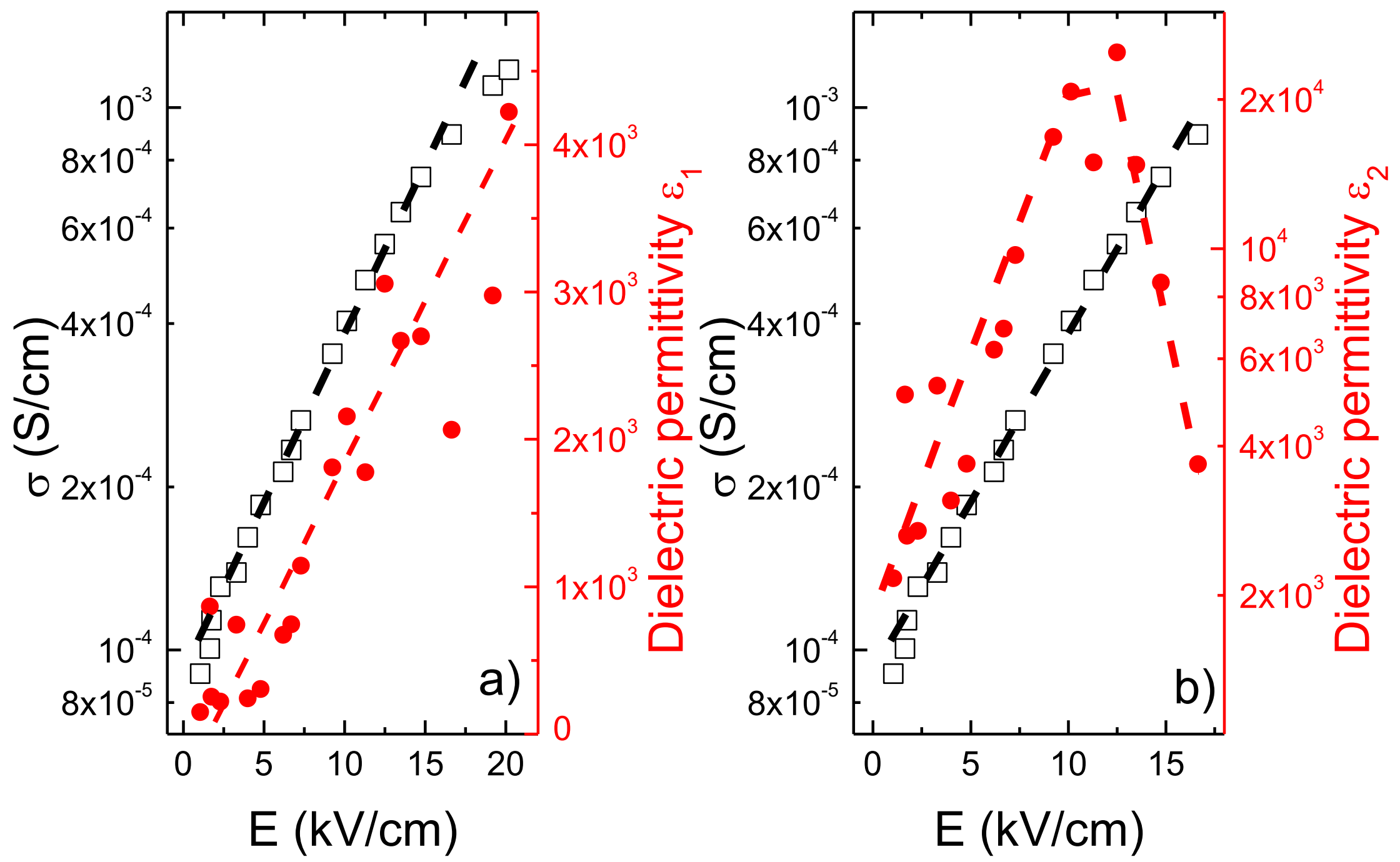

**Figure 7.** Dependences of conductivity σ and dielectric permittivity ε determined from the current decay curves vs electric field strength. Part **(a)** shows σ in comparison with $\varepsilon_1$, part **(b)** shows σ in comparison with $\varepsilon_2$.

## 4. Conclusion

The barium titanate grown by the Spark Plasma Sintering method with additional contamination by carbon possesses a colossal effective dielectric permittivity, up to several million at 120 °C. Its permittivity remains extremely high at low temperatures down to 77 K. In the temperature range 77-200 K, which corresponds to the rhombohedral phase in the single crystalline $BaTiO_3$, the dielectric permittivity and electric conductivity strongly depend on the carbon content. The more is the carbon content, the more permittivity and electric conductivity. Along with it the conductivity vs temperature dependence obeys in this temperature range the Mott law for the variable range hopping conduction. The conductivity has an electronic kind as obtained from measurement of thermo-emf. The numerical estimates based on measurements both of permittivity and conductivity show that the localization radius of electrons on the localized centers is about 3 nm. The dielectric permittivity determining the localization radius more corresponds to its high frequency measured magnitude. Along with this, there remains still poorly understood how one must consider the variation of the dielectric permittivity with temperature in the known models and theories of the hopping conduction.

Measurements of the current-voltage characteristics at fixed temperatures in the low temperature range showed them to obey the dependence of $ln\sigma \sim E$ in the range of moderately strong electric fields, which is also adherent to the known theory of the VRH conduction. The magnitude of the dielectric permittivity in this case also occurred to depend noticeably on the electric field, that does not find so far complete understanding within the frames of known theory of hopping conduction for a homogeneous semiconductor. In the case under consideration, we have a combined system consisting of nanosized core-shell $BaTiO_3$ grains with their inner hopping electric transport, their surface coated to some extent by carbon atoms and defects participating in hopping transfer of charge and ionic transport in the intergrain space. The decay current curves with two-exponential decay on applying rectangular high voltage pulses evidence of at least two component polarization processes.

## Appendix

**Table 1.** Dimensions of the samples, nominal carbon content and comparative properties

| Sample # | Contact surface dimensions (cm) | Sample thickness (cm) | Nominal carbon content (wt. %) | Comparative conductivity* | Lateral correlation length of carbon regions (nm) |
|---|---|---|---|---|---|
| S#1 | 0.33x0.2 | 0.2 | 0.5 | highest | 18 - 26 |
| S#2 | 0.254x0.177 | 0.162 | 0.5 | average | 23 - 29 |
| S#3 | 0.242x0.215 | 0.211 | 1 | lowest | 19 - 21 |
| S#4 | 0.6x0.58 | 0.158 | 0 | “absent”** | N/A |
| S#2.1 | 0.31x0.275 | 0.05 | 0.5 | average | 23 - 29 |
| S#3.1 | 0.25x0.2 | 0.04 | 1 | lowest | 19 - 21 |

[*] We would like to underline that the counterintuitive discrepancy between the nominal carbon addition and comparative carbon indicators, based on the electrical conductivity measurements (such as shown in Fig. 2) may originate from several reasons, and first, such as the inhomogeneous spatial distribution of carbon inclusions, which determines the electrical connectivity of carbon-rich regions, and their aggregation/clusterization.

[**] The resistivity of sample S#4 was several orders of magnitude higher than even that of sample S#3 and quickly went beyond measurable values as the temperature decreased in the VRH range.

**Author's contribution.** The idea of this research belongs to V.V.V and O.S.P. Electrophysical measurements and analysis of experimental results were carried out by O.S.P., T.O.K., V.V.V., D.O.S. and O.V.B. S.E.I. synthesized the ceramic samples. I.M.D. performed and analyzed Raman results. V.V.V. wrote the manuscript. A.N.M. and V.M.P. discussed the results and improved the text of the article.

**Acknowledgements.** The authors are indebted to Prof. Boris I. Shklovskii for helpful discussion. The work of O.S.P., D.O.S., A.N.M. is funded by the National Science Foundation of Ukraine (grants No. 2023.03/0132 “Multiple degenerated metastable states of spontaneous polarization in nanoferroics: theory, experiment and prospects for digital nanoelectronics”). The work of V.V.V. and V.M.P. is funded by the target program of the NAS of Ukraine, project No. 5.8/26-P "Energy-saving and environmentally friendly nanoscale ferroics for the development of sensors, nanoelectronics and spintronics. The research of T.O.K. was supported by the Ministry of Education and Science of Ukraine within the framework of the PhD Research Projects Competition. S.E.I. and A.N.M. also acknowledge the NATO Science for Peace and Security Programme under grant SPS G5980 “FRAPCOM” and Ministry of Education and Science of Ukraine (grant M/14-2026) for sponsoring nanoparticles preparation and characterization.